\documentstyle[11pt,aaspp4]{article}

\slugcomment{To appear in the Astrophysical Journal}

\lefthead{Bjorkman \& Wood}
\righthead{Monte Carlo Radiative Equilibrium}

\begin{document}

\title{Radiative Equilibrium and Temperature Correction \\
            in Monte Carlo Radiation Transfer}

\author{J.E.~Bjorkman}
\affil{Ritter Observatory, Department of Physics and Astronomy, 
University of Toledo, Toledo, OH 43606-3390; jon@astro.utoledo.edu}

\and

\author{Kenneth Wood}
\affil{Harvard-Smithsonian Center for Astrophysics,
60 Garden Street, Cambridge, MA 02138; kwood@cfa.harvard.edu}


\begin{abstract}

We describe a general radiative equilibrium and temperature correction
procedure for use in Monte Carlo radiation transfer codes with sources
of temperature-independent opacity, such as astrophysical dust.  The 
technique utilizes the fact that Monte Carlo simulations track 
individual photon packets, so we may easily determine where their energy is
absorbed.  When a packet is absorbed, it heats a particular cell within
the envelope, raising its temperature.  To enforce radiative
equilibrium, the absorbed packet is immediately re-emitted.  To correct
the cell temperature, the frequency of the re-emitted packet is chosen
so that it corrects the temperature of the spectrum previously emitted
by the cell.  The re-emitted packet then continues being scattered,
absorbed, and re-emitted until it finally escapes from the envelope.
As the simulation runs, the envelope heats up, and the emergent
spectral energy distribution (SED) relaxes to its equilibrium value,
{\it without iteration}.  This implies that the equilibrium temperature
calculation requires no more computation time than the SED calculation
of an equivalent pure scattering model with fixed temperature.  In
addition to avoiding iteration, our method conserves energy exactly,
because all injected photon packets eventually escape.  Furthermore,
individual packets transport energy across the entire system because
they are never destroyed.  This long-range communication, coupled with
the lack of iteration, implies that our method does not suffer the
convergence problems commonly associated with $\Lambda$-iteration.  To 
verify our temperature correction procedure, we compare our results to 
standard benchmark tests, and finally we present the results of 
simulations for two-dimensional axisymmetric density structures.

\end{abstract}


\section{Introduction}

There is an ever increasing wealth of observational evidence indicating
the non-sphericity of almost every type of astronomical object (e.g.,
extended circumstellar environments, novae shells, planetary nebulae,
galaxies, and AGNs).  To accurately interpret this data, detailed two-
and three-dimensional radiation transfer techniques are required.  With
the availability of fast workstations, many researchers are turning to
Monte Carlo techniques to produce model images and spectra for the
asymmetric objects they are investigating.  In Monte Carlo radiation
transfer simulations, packets of energy or ``photons'' are followed as
they are scattered and absorbed within a prescribed medium.  One of the
features of this technique is that the locations of the packets are
known when they are absorbed, so we can determine where their energy is
deposited.  This energy heats the medium, and to conserve radiative
equilibrium, the absorbed energy must be reradiated at other
wavelengths, depending on the opacity sources present.  Tracking these
photon packets, while enforcing radiative equilibrium, permits the
calculation of both the temperature structure and emergent spectral
energy distribution (SED) of the envelope.  The ability of Monte Carlo
techniques to easily follow the transfer of radiation through complex
geometries makes them very attractive methods for determining the
temperature structure within non-spherical environments --- a task
which is very difficult with traditional ray tracing techniques.

Previous work on this problem for spherical geometries includes the
approximate solutions by Scoville \& Kwan (1976), who ignored
scattering, Leung (1976), and diffusion approximations by Yorke
(1980).  The spherically symmetric problem has been solved exactly by
Rowan-Robinson (1980), Wolfire \& Cassinelli (1986), and Ivezi{\'c} \&
Elitzur (1997), who used a scaling technique.  Extensions of the exact
solution to two dimensions have been performed by Efstathiou \&
Rowan-Robinson (1990, 1991), while approximate two-dimensional models
have been presented by Sonnhalter, Preibisch, \& Yorke (1995) and
Men'shchikov \& Henning (1997).

Radiative equilibrium calculations using the Monte Carlo technique have
been presented by Lefevre, Bergeat, \& Daniel (1982); Lefevre, Daniel,
\& Bergeat (1983); Wolf, Henning, \& Secklum (1999); and Lucy (1999).
Most of these authors (Lucy being exceptional) use a technique in which
stellar and envelope photon packets are emitted separately.  The number of
envelope packets to be emitted is determined by the envelope
temperature, while the envelope temperature is determined by the number 
of absorbed packets.  Consequently these techniques require iteration,
usually using the absorbed stellar photons to provide an initial guess
for the envelope temperature.  The iteration proceeds until the
envelope temperatures converge.  Note that the stellar luminosity is
not automatically conserved during the simulation; only after the
temperatures converge is the luminosity approximately conserved.

In contrast, Lucy adopts a strategy in which absorbed photon packets are
immediately re-emitted, using a frequency distribution set by the
current envelope temperature.  Although the frequency distribution of
the reprocessed photons is incorrect (until the temperatures have
converged), his method automatically enforces local radiative
equilibrium and implicitly conserves the stellar luminosity.  The
insight of Lucy's method is that conservation of the stellar luminosity
is more important than the spectral energy distribution when
calculating the radiative equilibrium temperatures.  Nonetheless, this
method requires iteration.

The primary problem faced by Lucy's method is the incorrect frequency
distribution of the re-emitted photons.  In this paper we develop an
adaptive Monte Carlo technique that corrects the frequency distribution
of the re-emitted photons.  Essentially, our method relaxes to the
correct frequency and temperature distribution.  Furthermore it
requires no iteration as long as the opacity is independent of
temperature.  Such is the case for astrophysical dust.  In Section~2,
we describe the temperature correction algorithm.  We compare the
results of our code with a spherically symmetric code in Section~3, and
in Section~4 we present results for two dimensional axisymmetric
density structures.

\section{Monte Carlo Radiative Equilibrium} 
We wish to develop a method to calculate the temperature distribution
throughout an extended dusty environment for use with Monte Carlo
simulations of the radiation transfer.  The radiation transfer
technique we employ has been described in detail in other papers:  Code
\& Whitney (1995); Whitney \& Hartmann (1992, 1993); Wood et al.
(1996), so we only summarize it here.  The basic idea is to divide the
luminosity of the radiation source into equal-energy, monochromatic
``photon packets'' that are emitted stochastically
by the source.  These packets are followed to random interaction
locations, determined by the optical depth, where they are either
scattered or absorbed with a probability given by the albedo.  If the
packet is scattered, a random scattering angle is obtained from the
scattering phase function (differential cross section).  If instead the
packet is absorbed, its energy is added to the envelope, raising the
local temperature.  To conserve energy and enforce radiative
equilibrium, the packet is re-emitted immediately at a new frequency
determined by the envelope temperature.  These re-emitted photons
comprise the diffuse radiation field.  After either scattering or
absorption plus reemission, the photon packet continues to a new interaction
location.  This process is repeated until all the packets escape the
dusty environment, whereupon they are placed into frequency and
direction-of-observation bins that provide the emergent spectral energy
distribution.  Since all the injected packets eventually escape (either
by scattering or absorption followed by reemission), this method
implicitly conserves total energy.  Furthermore it automatically
includes the diffuse radiation field when calculating both the
temperature structure and the emergent spectral energy distribution.

We now describe in detail how we calculate the temperature structure
and SEDs of dusty environments illuminated by a radiation source.  This
radiation can come from any astrophysical source, either internal or
external, point-like or extended.

\subsection{Radiative Equilibrium Temperature} 

Initially we divide the source luminosity, $L$, into $N_\gamma$ photon
packets emitted over a time interval $\Delta t$.  Each photon packet
has the same energy, $E_\gamma$, so
\begin{equation}
                           E_\gamma = L \Delta t / N_\gamma \; .
\label{eq:Ephot}
\end{equation} 
Note that the number of physical photons in each packet is 
frequency-dependent.

When the monochromatic photon packet is injected into the envelope, it
will be assigned a random frequency chosen from the spectral energy
distribution of the source.  This frequency determines the dust
absorptive opacity, $\kappa_\nu$, and scattering opacity, $\sigma_\nu$
(both per mass), as well as the scattering parameters for the ensuing
random walk of the packet through the envelope.  The envelope is divided 
into spatial grid cells with volume $V_i$, where $i$ is the cell index.  
As we inject source photon packets, we maintain a running total, $N_i$, 
of how many packets are absorbed in each grid cell.  Whenever a packet 
is absorbed in a grid
cell, we deposit its energy in the cell and recalculate the cell's
temperature.  The total energy absorbed in the cell
\begin{equation}
                       E_i^{\rm abs} = N_i E_\gamma \; .
\label{eq:Eabs} 
\end{equation} 

We assume that the dust particles are in local thermodynamic
equilibrium (LTE), and for simplicity we adopt a single temperature for
the dust grains.  Note that although we use dust for the continuous
opacity source, we could replace the dust by any continuous LTE opacity
source that is independent of temperature.  In radiative equilibrium,
the absorbed energy, $E_i^{\rm abs}$, must be reradiated.  The thermal
emissivity of the dust $j_\nu = \kappa_\nu \rho B_\nu(T)$, where
$B_\nu(T)$ is the Planck function at temperature $T$, so the emitted 
energy is
\begin{eqnarray}
 E_i^{\rm em} &=& 4\pi\Delta t \int dV_i \int \rho \kappa_\nu B_\nu(T) \,d\nu \cr
              &=& 4\pi\Delta t \int \kappa_P(T) B(T) \rho    \,dV_i\;,
\end{eqnarray} 
where $\kappa_P=\int\kappa_\nu B_\nu\,d\nu / B$ is the Planck mean
opacity, and $B = \sigma T^4 / \pi$ is the frequency integrated Planck
function.  If we adopt a temperature that is constant throughout the
grid cell, $T_i$, then
\begin{equation}
             E_i^{\rm em} = 4\pi\Delta t \kappa_P(T_i) B(T_i) m_i \;,
\label{eq:Eem} 
\end{equation} 
where $m_i$ is the mass of the cell.

Equating the absorbed (\ref{eq:Eabs}) and emitted (\ref{eq:Eem})
energies, we find that after absorbing $N_i$ packets, the dust
temperature is given by
\begin{equation}
    \sigma T_i^4= { {N_i L }\over {4 N_\gamma \kappa_P(T_i) m_i} } \; .
\label{eq:REtemp} 
\end{equation}
Note that the Planck mean opacity, $\kappa_P$, is a function of 
temperature, so 
equation~(\ref{eq:REtemp}) is actually an implicit equation for the 
temperature, which must be solved {\it every time a packet is absorbed}.  
Since this equation is solved so many times, an efficient algorithm is 
desirable.  Fortunately $\kappa_P$ is a slowly varying function of 
temperature, which implies a simple iterative algorithm may be used to
solve equation~(\ref{eq:REtemp}).  To do so efficiently, we pre-tabulate 
the Planck mean opacities for a large range of temperatures and evaluate 
$\kappa_P(T_i)$ by interpolation, using the temperature from the previous
guess.  After a few steps, we have the solution for $T_i$.  Note that 
because the dust opacity is temperature-independent, the product 
$\kappa_P \sigma T_i^4$, which is proportional to $\int\kappa_\nu B_\nu\,d\nu$,
increases monotonically with temperature.  
Consequently $T_i$ always increases when the cell absorbs an additional 
packet.

\subsection{Temperature Correction}

\begin{figure}
\epsscale{0.8}
\plotone{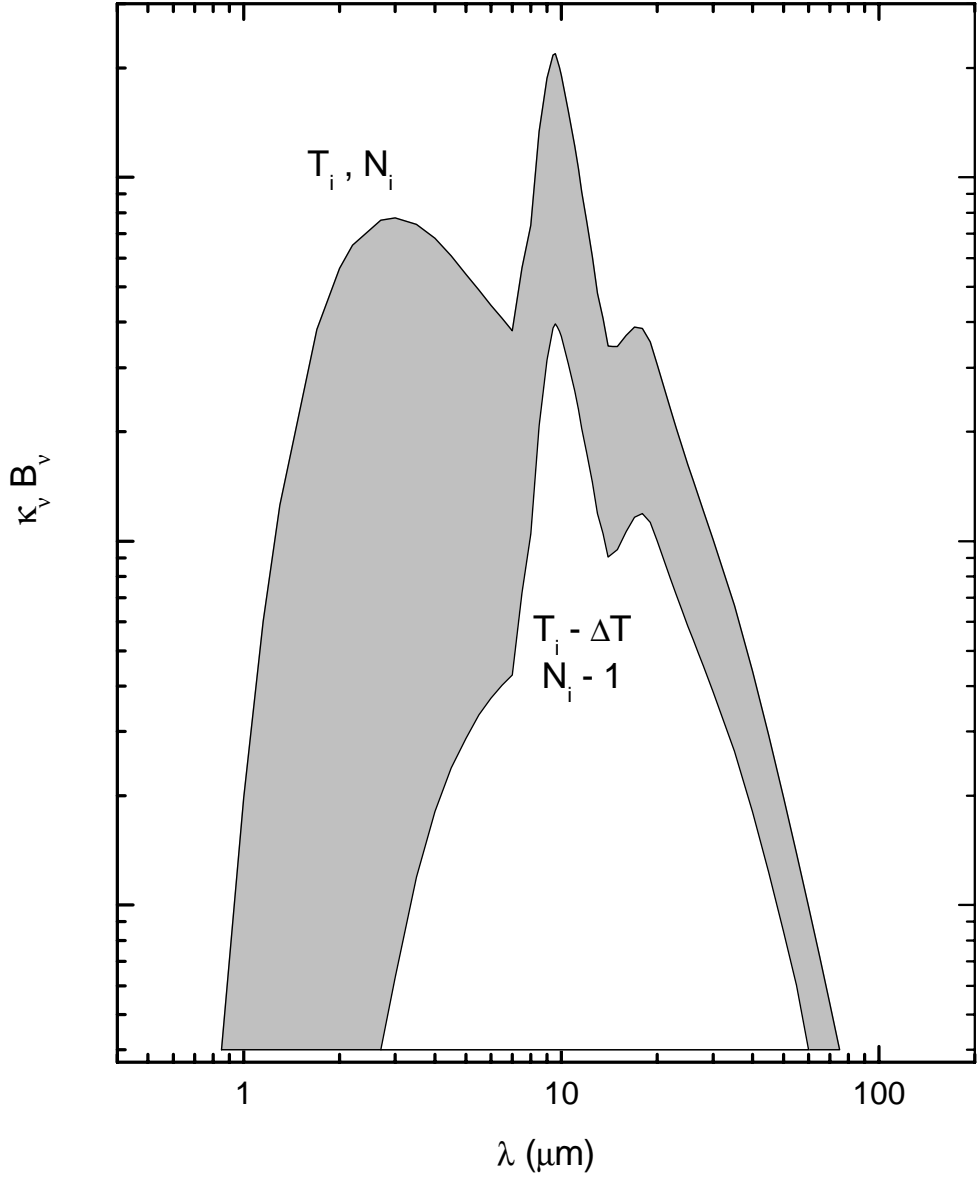}
\figurenum{1}
\caption[] {Temperature Correction Frequency Distribution.  Shown are
the dust emissivities, $j_\nu = \kappa_\nu B_\nu(T)$, prior to and after
the absorption of a single photon packet.  The spectrum of the previously 
emitted packets is given by the emissivity at the old cell temperature 
(bottom curve).  To correct the spectrum from the old temperature to the new 
temperature (upper curve), the photon packet should be re-emitted using the 
difference spectrum (shaded area).}
\label{fig:tempcorrect}
\end{figure}

Now that we know the temperature after absorbing an additional packet
within the cell, we must reradiate this energy so that the heating
always balances the cooling.  Prior to absorbing this packet, the cell
previously has emitted packets that carried away an energy
corresponding to the cell's previous emissivity $j^\prime_\nu =
\kappa_\nu B_\nu(T_i - \Delta T)$, where $\Delta T$ is the temperature
increase arising from the packet absorption.  Note that these previous
packets were emitted with an incorrect frequency distribution
corresponding to the previous temperature, $T_i-\Delta T$.  The total
energy that should be radiated now corresponds to $j_\nu$ at the new
temperature, $T_i$.  Thus the additional energy to be carried away is
given by
\begin{equation}
  \Delta j_\nu = j_\nu - j^\prime_\nu 
               = \kappa_\nu\left[B_\nu(T_i) - B_\nu(T_i -\Delta T)\right] \; ,
\label{eq:DeltaKappaB}
\end{equation}
which is the shaded area shown in Figure~\ref{fig:tempcorrect}.  As long 
as the packet energy $E_\gamma$ is not too large (this may be assured
by choosing a large enough number of photon packets, $N_\gamma$, to use for 
the simulation), the temperature change $\Delta T$ is small, so the 
temperature correction spectrum
\begin{equation}
       \Delta j_\nu \approx \kappa_\nu \Delta T {{dB_\nu} \over {dT}} \; .
\label{eq:Deltajnu}
\end{equation}
Note that $\Delta j_\nu$ is everywhere positive because $\Delta T > 0$, and 
the Planck function is a monotonically increasing function of temperature.  
Therefore to correct the previously emitted spectrum, we immediately re-emit 
the packet (to conserve energy), and we choose its frequency using the shape 
of $\Delta j_\nu$.  This procedure statistically reproduces $\Delta j_\nu$
for the distribution of the re-emitted packets.  Normalizing this 
distribution, we find the temperature correction probability distribution
\begin{equation}
       {{dP_i} \over {d\nu}} = {{\kappa_\nu}\over{K}} \left ({{dB_\nu} \over 
{dT}}\right )_{T=T_i} \; ,
\label{eq:Tcorrect} 
\end{equation}
where $dP_i /d\nu$ is the probability of re-emitting the packet between
frequencies $\nu$ and $\nu+d\nu$, and the normalization constant 
$K=\int_0^\infty \kappa_\nu (dB_\nu/dT)\,d\nu$.

\placefigure{fig:tempcorrect}

Now that the packet's frequency has changed, we change the opacity and
scattering parameters accordingly and continue with scattering,
absorption, temperature correction, and re-emission until all the photon 
packets finally escape from the system.  In principle we could also
account for back-warming of the source.  Whenever a packet hits the
source, the source must reradiate this new energy.  This will change
the temperature of the source, and the new source photons can be emitted
using a difference spectrum similar to equation (\ref{eq:Tcorrect}).

When we begin our calculation, no packets have been absorbed, so the
initial temperature is zero throughout the envelope.  This means that
the initial temperature change is not small as is required by 
equation~(\ref{eq:Deltajnu}).  One could use 
equation~(\ref{eq:DeltaKappaB}) to re-emit the first packet that is
absorbed; however, this is not necessary.  The number of packets 
producing this initial temperature change is small; it is of order the
number of spatial grid cells.  Furthermore, these packets generally are 
re-emitted at such long wavelengths that they are not observable.  
Consequently the error arising from using equation~(\ref{eq:Deltajnu}) 
to re-emit every packet is too small to be of importance.

As the simulation runs, the envelope starts at zero temperature. It
then heats up, and the radiation field ``relaxes'' to its final spectral 
shape.  The temperature correction procedure is simply a way of re-ordering 
the calculation (which frequencies are being used at a given moment) so that 
in the end, all the frequencies have been properly sampled.  Consequently,
after all the stellar photon packets have been transported, the final
envelope temperature is the correct radiative equilibrium temperature, 
and the emergent spectral energy distribution has the correct frequency
shape.  Furthermore, the photon re-emission automatically accounts for the
diffuse envelope emission.  Note that energy is necessarily conserved,
and there is no time-consuming iteration in our scheme; calculating the
radiative equilibrium temperature requires no more computational time
than an equivalent pure scattering calculation in which the temperature
structure is held fixed.  Similarly, there is no issue of convergence
in our method.  Unlike $\Lambda$-iteration, the photon packets carry 
energy over large distances throughout the envelope because they are never
destroyed, and of course there is no iteration at all.  After running 
$N_\gamma$ packets, we have the final answer.  The simulation does not 
continue running until some convergence criterion is met, and the only
source of error in the calculation is the statistical sampling error 
inherent in Monte Carlo simulations.

\section{Benchmark Verification}
To validate our method for determining the radiative equilibrium
temperatures and emergent fluxes, we compared our results against a set
of benchmark calculations recently developed by Ivezi{\' c} et
al.~(1997) for testing spherically symmetric dust radiative equilibrium
codes. The parameters listed by Ivezi{\' c} et al. enable us to exactly
reproduce the same set of physical conditions (i.e, input spectrum,
dust destruction radius, optical depth, opacity frequency distribution, 
and radial density structure).

For all benchmark tests, Ivezi{\' c} et al. used a point source star,
radiating with a black body spectrum whose temperature $T_\star =
2500$K.  The dust density distribution was a power law with radius,
\begin{equation}
     \rho = \rho_0 \left({{R_{\rm dust}}\over{r}}\right)^p \;,
\label{eq:rho}
\end{equation}
where
\begin{equation}
 \rho_0 =  {                   {\tau_\lambda} \over
    {(\kappa_\lambda+\sigma_\lambda)(1-R_{\rm dust}/R_{\rm max})}}
       \cases{ 1/R_{\rm max}  & ($p=0$),\cr
               1/R_{\rm dust} & ($p=2$).\cr}
\label{eq:rho0}
\end{equation}
The inner radius of the envelope is the dust destruction radius, 
$R_{\rm dust}$, the outer radius is $R_{\rm max}$, and the total radial 
optical depth is $\tau_\lambda$, specified at $\lambda = 1\mu {\rm m}$.  
The dust absorptive opacity, $\kappa_\nu$, and scattering opacity, 
$\sigma_\nu$, were taken to be
\begin{eqnarray}
{{\kappa_\nu}\over{(\kappa+\sigma)_{1\mu{\rm m}}}}  
          &=& 0.5 \cases{ 1            &($\lambda<1\,\mu{\rm m}$),\cr
      (1\mu{\rm m} / \lambda)\phantom{^4} &($\lambda>1\,\mu{\rm m}$),\cr} \cr 
{{\sigma_\nu}\over{(\kappa+\sigma)_{1\mu{\rm m}}}}  
          &=& 0.5 \cases{ 1            &($\lambda<1\,\mu{\rm m}$),\cr
      (1\mu{\rm m} / \lambda)^4 &($\lambda>1\,\mu{\rm m}$).\cr}
\end{eqnarray}
Since the total optical depth at $1\mu {\rm m}$ is independently
specified, these opacities have been normalized to that at $1\mu {\rm
m}$ for convenience. The wavelength-dependent scattering albedo is
given by $a = \sigma_\nu/(\kappa_\nu+\sigma_\nu)$, and the scattering
was assumed to be isotropic (note that in dust simulations, we would
normally use a non-isotropic phase function for the scattering).

In principle, the inner radius of the dust shell, $R_{\rm dust}$, is
determined by the dust condensation temperature, chosen by Ivezi{\' c} 
et al. to be $T_{\rm cond} = 800\,{\rm K}$.  However, we are only testing 
the temperature correction procedure, so we have not implemented a scheme 
to solve self consistently for the dust destruction radius. The values for 
$R_{\rm dust}$ were calculated instead using Ivezi{\' c} et al.'s eq.~(4) 
and data from their Table~1.  Finally, the outer radius of the dust shell
was set to be $R_{\rm max} = 10^3 R_{\rm dust}$.  The parameters
describing the various test simulations are summarized in 
Table~\ref{table:benchmarkparms}.

\placetable{table:benchmarkparms}

\begin{deluxetable}{lcccc}
\tablewidth{0pt}
\tablecaption{Spherical Models}
\tablehead{ \colhead {$p$} & \colhead{$\tau_{1\mu{\rm m}}$} & 
\colhead {$R_{\rm dust}/R_\star$} }
\startdata
0 & 1   & 8.44 \nl
0 & 10  & 8.46 \nl
0 & 100 & 8.60 \nl
2 & 1   & 9.11 \nl
2 & 10  & 11.37 \nl
2 & 100 & 17.67 \nl
\enddata
\label{table:benchmarkparms}
\end{deluxetable}

\placetable{table:ellipsoidparms}

\begin{deluxetable}{lcccc}
\tablewidth{0pt}
\tablecaption{Ellipsoidal Models}
\tablehead{ \colhead {$\rho_{\rm eq}/\rho_{\rm pole}$} & 
\colhead {$R_{\rm dust}/R_\star$} & \colhead{$\tau^{\rm eq}_V$} & 
\colhead{$\tau^{\rm pole}_V$} }
\startdata
1000   & 10   & 200   & 0.2 \nl
1000   & 10   & 20    & 0.02 \nl
10     & 10   & 200   & 20  \nl
10     & 10   & 20    & 2 \nl
\enddata
\label{table:ellipsoidparms}
\end{deluxetable}

To begin the simulation, we release stellar photon packets with a black body
frequency distribution, given by the normalized Planck function
\begin{equation}
b_\nu (x) = {15\over{\pi^4}} { {x^3}\over{{\rm e}^x - 1} } \; ,
\end{equation}
where $x=h\nu/kT_\star$.  A particularly simple method for sampling the
black body distribution is given by Carter \& Cashwell (1975).  Since
this reference is somewhat obscure and difficult to obtain, we summarize
the method here.  First, generate a uniform random number, $\xi_0$, in 
the range 0 to 1, and determine the minimum value of $l$,
$l_{\rm min}$, that satisfies the condition
\begin{equation}
\sum_{i=1}^{l} i^{-4} \ge {{\pi^4}\over{90}} \xi_0 \; .
\end{equation}
Next obtain four additional uniform random numbers (in the range 0 to 1),
$\xi_1$, $\xi_2$, $\xi_3$, and $\xi_4$.  Finally, the packet 
frequency is given by
\begin{equation}
x = -\ln{(\xi_1\xi_2\xi_3\xi_4)}/l_{\rm min}\; .
\end{equation}

After emitting these packets from the star, we track them through the
envelope.  To determine the envelope temperature, we must count how many 
packets are absorbed in each grid cell. Since the envelope is spherically 
symmetric, we employ a set of spherical shells for our grid.  To obtain 
the best Poisson error statistics, we should ideally construct the grid 
positions so that equal numbers of packets are absorbed in each cell.  
Since the probability of absorbing a photon packet is proportional to the 
optical depth, we choose equal radial optical depth grid locations,
\begin{equation}
{{r_i} \over {R_{\rm dust}} } = \cases{ i \Delta r + 1  & ($p=0$),\cr
                                        N_f / (N_f - i) & ($p=2$),\cr}
\end{equation}
where $N_r=200$ is the total number of cells we used, 
$\Delta r = (R_{\rm max}/R_{\rm dust}-1)/N_r$, and 
$N_f = N_r/(1-R_{\rm dust}/R_{\rm max})$.  Integrating the density,
eq.~(\ref{eq:rho}), over the cell volume to obtain the mass, we find
from eq.~(\ref{eq:REtemp}) that the temperature in each grid cell is 
given by
\begin{equation}
T_i^4 = T_\star^4 \cases{ { {N_i N_r (R_\star/R_{\rm dust})^2
        [(\kappa_{1\mu{\rm m}}+\sigma_{1\mu{\rm m}})/\kappa_P(T_i)]} \over 
                { 4 N_\gamma \tau_{1\mu{\rm m}}   
        [(i^2-i+1/3)\Delta r^2+(2i-1)\Delta r+1 ] }} & ($p=0$),\cr
                {{ N_i(N_f-i)(N_f-i+1)(R_\star/R_{\rm dust})^2 
                (1-R_{\rm dust}/R_{\rm max})} \over
        {4 N_\gamma \tau_{1\mu{\rm m}} N_f [\kappa_P(T_i) / 
        (\kappa_{1\mu{\rm m}}+\sigma_{1\mu{\rm m}})]}} & ($p=2$),\cr} 
\end{equation}
where we have used $L=4\pi R_\star^2\sigma T_\star^4$ for the stellar 
luminosity.  

We then proceed with the radiation transfer, temperature calculation, and 
reemission as described in Section~2 until all packets exit the envelope.  
When the packets escape, they are placed into $N_\nu=1024$ uniform frequency 
bins,
\begin{equation}
     \nu_k = \nu_{\rm max} {k \over N_\nu} \; ,
\label{eq:freqbins}
\end{equation}
where $h \nu_{\rm max} = 16 k T_\star$.  The width of each bin $\Delta \nu
= \nu_{\rm max} / N_\nu$.  Since the envelope is spherically symmetric, the 
observed flux $F_{\nu_k} = N_k E_\gamma / 4 \pi d^2 \Delta t \Delta \nu$, 
where $N_k$ is the number of packets in the $k^{\rm th}$ frequency bin, and 
$d$ is the observer's distance from the star.  Normalizing to the total 
flux, $F=L/4 \pi d^2$, the SED is given
by
\begin{equation}
  \left({{\nu F_\nu} \over {F}}\right)_k = (k-1/2) {N_k \over N_\gamma} \; .
\end{equation}
The factor $(k-1/2)$ arises from using the frequency at the center of the bin.

\begin{figure}
\epsscale{0.8}
\plotone{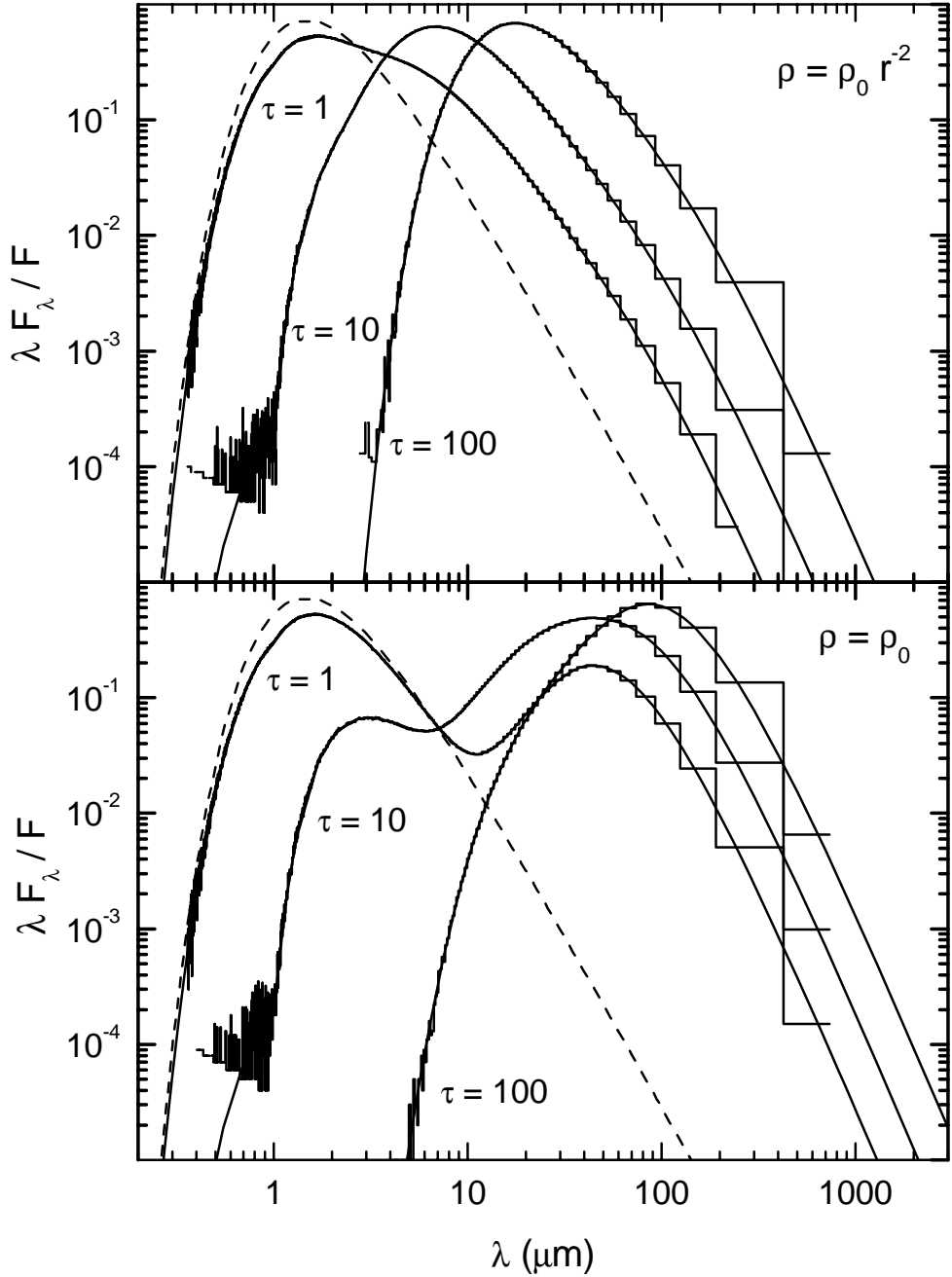}
\figurenum{2a}
\caption[] {Spherical Model SEDs.  Shown are the comparisons of our 
spherically symmetric model fluxes (histogram) with those from DUSTY 
(solid curves).  The top panel compares the $1/r^2$ density models, while
the bottom panel compares the constant density models.  The optical
depths at 1$\mu$m are as indicated.  The incident stellar spectrum is 
shown by the dashed curve.}
\label{fig:benchmarkSED}
\end{figure}

\begin{figure}
\epsscale{0.8}
\plotone{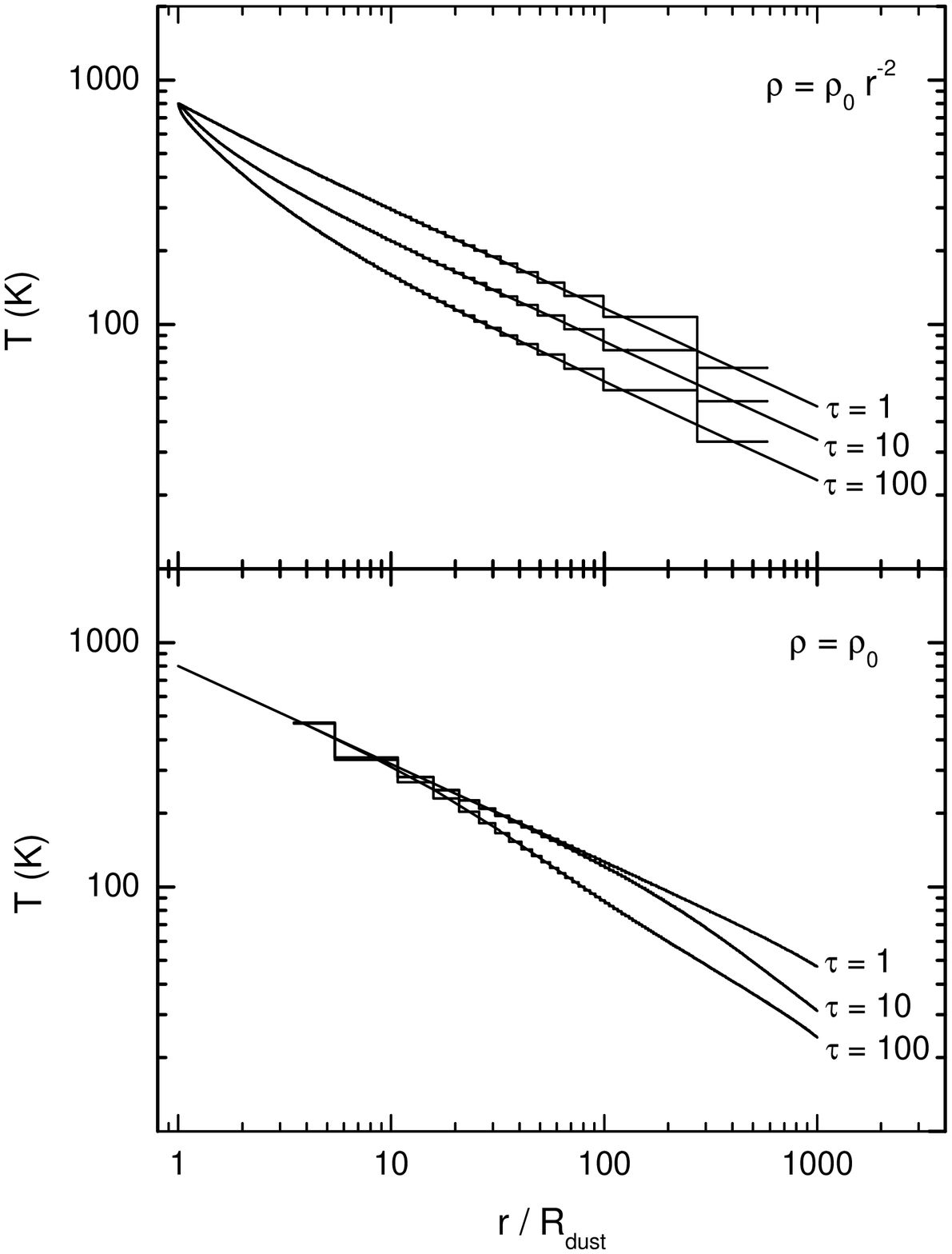}
\figurenum{2b}
\caption[] {Spherical Model Temperatures.  Shown are the comparisons of our 
spherically symmetric model temperatures (histogram) with those from DUSTY 
(solid curves).  The top panel compares the $1/r^2$ density models, while
the bottom panel compares the constant density models.  The optical depths 
at 1$\mu$m are as indicated.}
\label{fig:benchmarkTemp}
\end{figure}

\placefigure{fig:benchmarkSED}
\placefigure{fig:benchmarkTemp}
\notetoeditor{Please place Figs.~\ref{fig:benchmarkSED} \& 
\ref{fig:benchmarkTemp} side by side.}

In Figures~\ref{fig:benchmarkSED} and \ref{fig:benchmarkTemp} we show
the results of our simulation compared with the output of one of the
codes tested by Ivezi{\'c} et al.  This code, called DUSTY, is publicly
available and is described in Ivezi{\'c} \& Elitzur (1997).  We see
that our Monte Carlo calculations reproduce both the correct
temperature structure and SED.  Note the large errors at the longest
and shortest wavelengths in the Monte Carlo calculations.  At these
wavelengths, the number of emerging packets is small, resulting in
large errors (the relative error in each flux bin equals
$1/\sqrt{N_k}$, owing to the Poisson sampling statistics inherent in
Monte Carlo simulations).  The excellent agreement of the comparisons
shown in Figures~2a and 2b (the differences are smaller than
the numerical error of the DUSTY calculations) demonstrates the
validity of our temperature correction procedure described in
Section~2.

Now that we have verified our our basic radiative equilibrium algorithm, 
we can proceed to investigate the temperature structure and SEDs of other 
geometries.  Owing to the inherently three-dimension nature of Monte Carlo 
simulations (even our 1-D spherically symmetric code internally tracks the 
photon packets in three dimensions), our code is readily modified for arbitrary 
geometries.  We now show the results of an application to axisymmetric 
circumstellar environments.

\section{Axisymmetric 2-D Calculations}
For the purpose of this illustrative simulation, we adopt a stellar
blackbody with $T_\star=3500\ {\rm K}$, envelope inner radius 
$R_{\rm dust}=10R_\star$, and a simple ellipsoidal parameterization for 
the circumstellar density. The isodensity contours are elliptical with 
$a/b$ being the ratio of the semi-major to semi-minor axis.  The 
density is given by
\begin{equation}
\rho = \rho_0 { {(R_{\rm dust}/r)^2} \over {1+f^2\cos^2\theta} } \; ,
\end{equation}
where $\rho_0$ is given by eq.~(\ref{eq:rho0}; $p=2$ case), $\theta$ is 
the polar angle, and the ``flattening factor''
\begin{equation}
                       f=\sqrt{(a/b)^2-1} \; .
\end{equation}
Note that the equatorial to polar density ratio at a given radius is 
$\rho_{\rm eq}/\rho_{\rm pole} = (a/b)^2$.

As before, we divide the circumstellar environment into cells with 
$N_r = 200$ radial and $N_\mu = 20$ latitudinal grid points.  Note that
the envelope is symmetric about the equator, so we combine the cells
below the equator with their counterparts above the equator.  This is
automatically accomplished by using $\mu_j = \cos(\theta_j)$ for the
latitudinal grid point coordinate.  Spacing the grid so that the 
radial and latitudinal optical depths are the same for each cell, we
find
\begin{eqnarray}
    r_i &=& {{N_f} \over {N_f - i}} R_{\rm dust} \; , \cr  
  \mu_j &=& {{1} \over {\sqrt{1+(a/b)^2\tan^2(\pi j/2N_\mu)}}} \; .
\end{eqnarray}
With these cell coordinates, equation~(\ref{eq:REtemp}) for the 
temperature in cell $(i,j)$ becomes
\begin{equation}
T_{i,j}^4 = T^4_\star {{ N_{i,j}(N_f-i)(N_f-i+1)(R_\star/R_{\rm dust})^2 
            (1-R_{\rm dust}/R_{\rm max})f} \over
           {4 N_\gamma \tau^{\rm eq}_V N_f 
          [\kappa_P(T_{i,j}) / (\kappa_V+\sigma_V)]
              [\tan^{-1}(f\mu_{j-1})-\tan^{-1}(f\mu_{j})] }} \; ,
\end{equation}
where $N_{i,j}$ is the number of packets absorbed in the cell, and we have
chosen $\lambda = 5500$\AA\ ($V$-band) for our equatorial radial optical 
depth parameter, $\tau^{\rm eq}_V$.

\begin{figure}
\epsscale{0.8}
\plotone{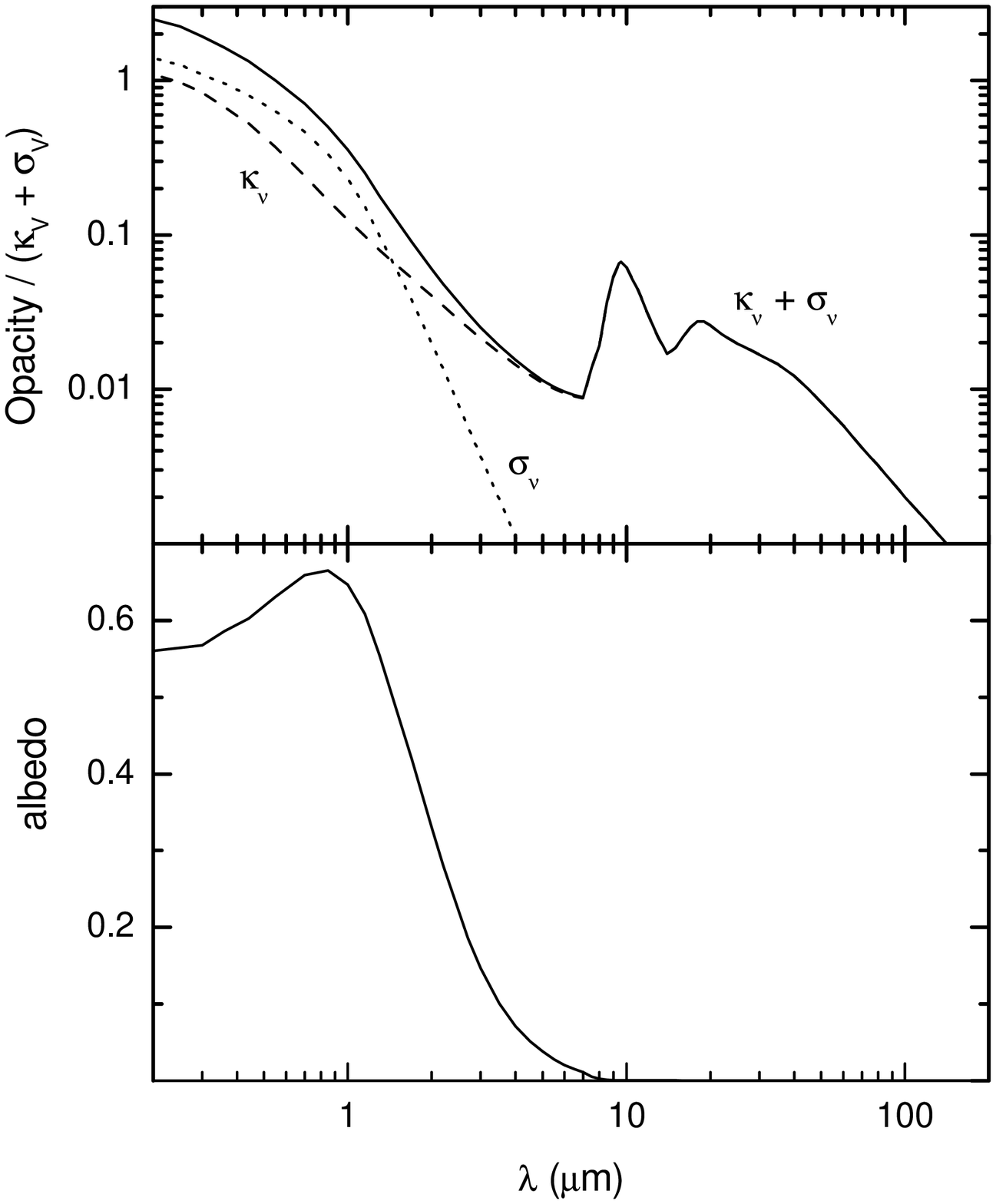}
\figurenum{3}
\caption[] {Dust Opacity.  The normalized absorptive (dashed line), 
scattering (dotted line), and total (solid line) opacities for the 2-D 
simulations are shown in the top panel.  The bottom panel shows the the 
corresponding albedo.}
\label{fig:dustopac}
\end{figure}

For the dust opacity, we adopt a standard MRN interstellar grain
mixture (Mathis, Rumpl, \& Nordsieck 1977), using optical constants
from Draine \& Lee (1984).  Figure~\ref{fig:dustopac} shows the opacity 
and albedo in graphical form.  The data for this figure is available
in tabular 
form\footnote[1]{ftp://gradj.pa.uky.edu/dusty/distribution/ism-stnd.dat} 
from the DUSTY Web site, http://www.pa.uky.edu/{$\sim$}moshe/dusty.  Note 
the prominent silicate absorption features at $10\mu$m and
$18\mu$m.  For the current demonstration, we assume that the scattering
is isotropic, but we can easily accommodate any phase function,
analytic or tabulated when calculating the emergent SED.

\placefigure{fig:dustopac}

Unlike the spherically symmetric benchmarks, the SED now depends on 
viewing angle, so we must bin the escaping packets both in direction 
and frequency.  Since the envelope is axisymmetric, the observed flux 
only depends on the inclination angle, $i$, of the envelope symmetry
axis (i.e., the stellar rotation axis).  To obtain approximately equal 
numbers of escaping packets, we choose $N_{\rm inc} = 10$ direction bins, 
with equal solid angles, $\Delta \Omega = 4 \pi / N_{\rm inc}$.  The
escaping directions (inclination angles), $i_l$, for these bins are
given by
\begin{equation}
   \mu^{\rm esc}_l = l/N_{\rm inc} \; ,
\end{equation}
where $\mu^{\rm esc}_l = \cos(i_l)$.  In addition to the direction bins,
we use the frequency bins given by eq.~(\ref{eq:freqbins}).

Now that we have specified both the direction and frequency bins, the 
observed flux is $F_{\nu_{k,l}} = N_{k,l} E_\gamma / \Delta \Omega d^2 
\Delta t \Delta \nu$, where $N_{k,l}$ is the number of escaping packets 
in the $(k,l)$ bin.  Normalizing to the bolometric flux F gives the 
emergent SED,
\begin{equation}
  \left({{\nu F_\nu} \over {F}}\right)_{k,l} = 
                    {(k-1/2){N_{k,l} N_{\rm inc}} \over {N_\gamma}} \; .
\end{equation}

We now choose to investigate the SEDs produced by two density
structures with different degrees of flattening.  The first has a
density ratio $\rho_{\rm eq} / \rho_{\rm pole} = 1000$ to represent a
disk-like structure.  The second has a density ratio $\rho_{\rm eq} /
\rho_{\rm pole} = 10$, which is mildly oblate, representative of an
infalling protostellar envelope.  For each density structure, we
perform optically thick (in the mid-IR) and optically thin
calculations.  The optically thick calculations have an equatorial
$V$-band optical depth $\tau^{\rm eq}_V=200$, while the optically thin
calculations have $\tau^{\rm eq}_V=20$.
Table~\ref{table:ellipsoidparms} summarizes these density structures.
For comparison, we also have performed calculations for $p=2$
spherically symmetric models containing the same total mass as our disk
and envelope densities.  Keeping the same inner and outer radii, the
optical depth for the equivalent spherical model is
\begin{equation}
\tau^{\rm sp}_V = \tau^{\rm eq}_V { {\tan^{-1} f} \over f }\; .
\end{equation}

\begin{figure}
\epsscale{0.8}
\plotone{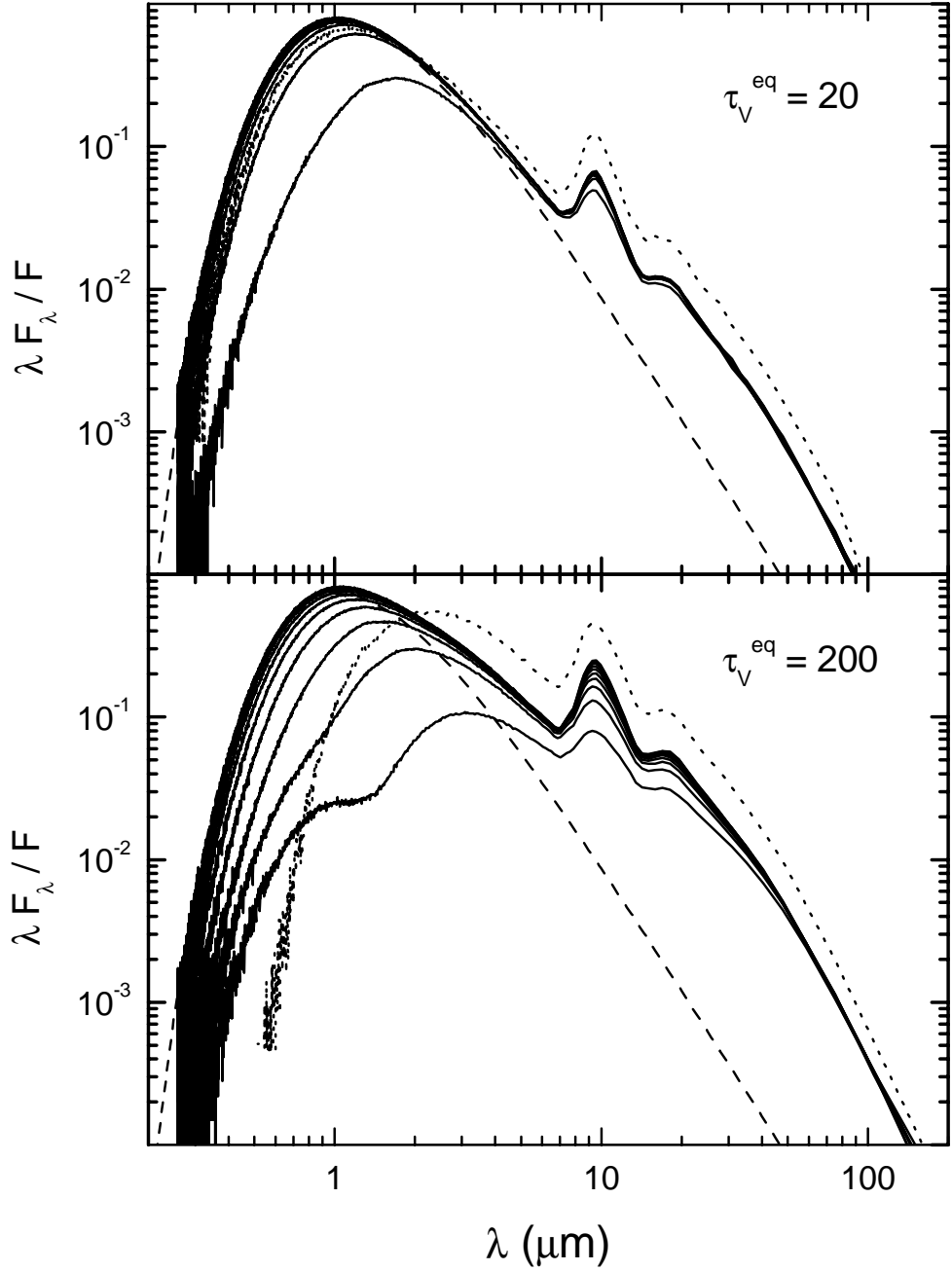}
\figurenum{4a}
\caption[] {Disk Model SEDs.  The normalized emergent fluxes (solid lines) 
are shown as a function of viewing angle (10 inclinations, evenly spaced 
in $\cos i$) for the disk-like circumstellar density.  The lowest curve 
corresponds to an almost edge-on view ($\cos i=0.05$), while the highest
curve corresponds to an almost pole-on view ($\cos i=0.95$).  The dotted 
curves are for a spherically symmetric simulation, having the same total 
circumstellar mass.  The incident stellar spectrum is shown by the dashed 
curve.}
\label{fig:diskSED}
\end{figure}

\begin{figure}
\epsscale{0.8}
\plotone{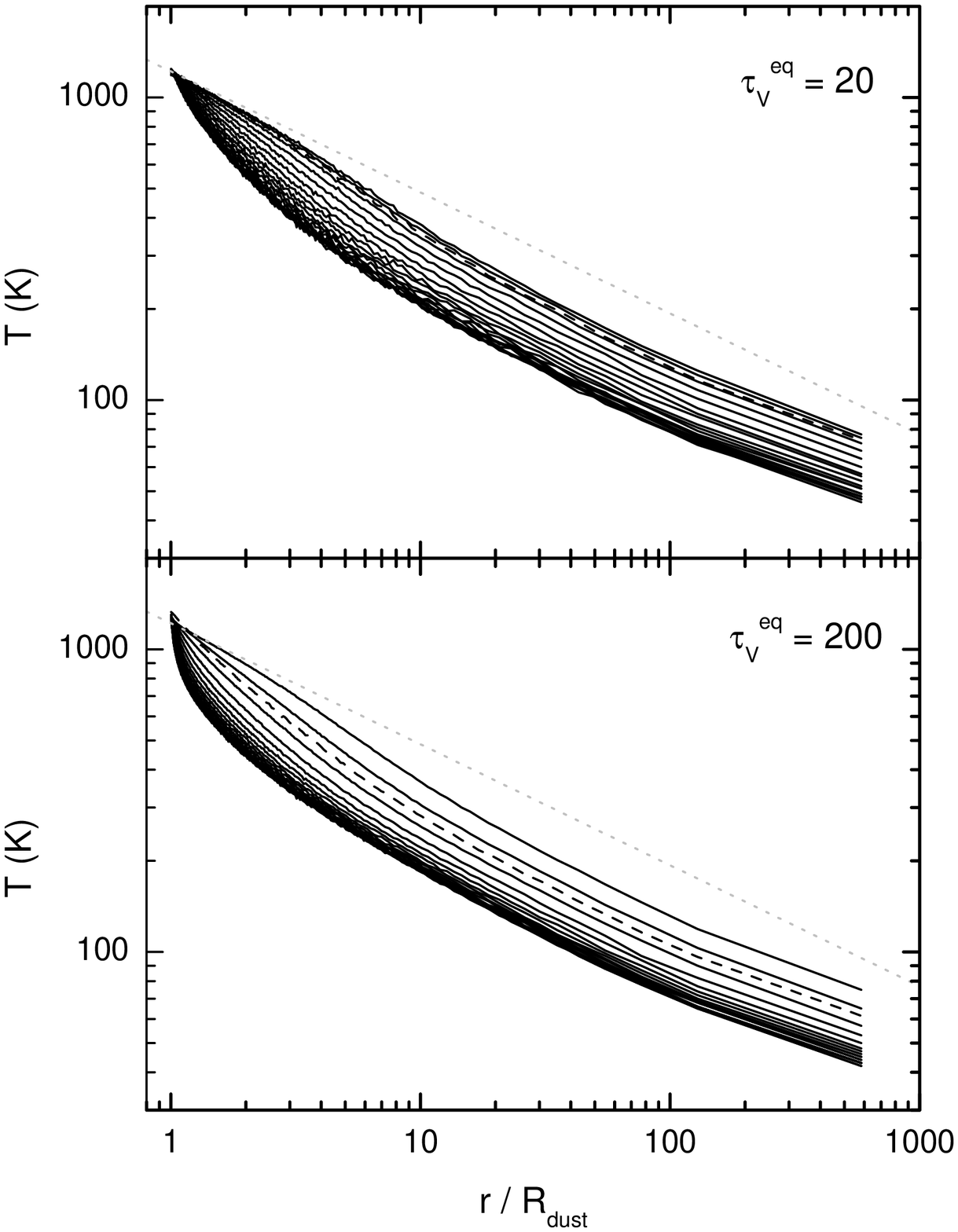}
\figurenum{4b}
\caption[]{Disk Model Temperatures.  The temperatures (solid lines) are 
shown as a function of polar angle (20 angles, spaced as indicated in
the text).  The lowest curve is closest to the equatorial plane and the
highest curve is nearest to the polar axis.  The dashed curve is the
temperature for a spherically symmetric simulation, having the same total 
circumstellar mass, and the light dotted curve is a simple power law $T 
\propto r^{-0.4}$ for reference.}
\label{fig:diskTemp}
\end{figure}

\subsection{Disk Model}

Figure~\ref{fig:diskSED} shows the incident stellar spectrum and the
emergent SED as a function of viewing angle for the disk-like model,
as well as the result of the equivalent spherically symmetric
calculation.  For both disk simulations, when viewing the system
pole-on, we are looking through optically thin circumstellar dust (see
Table~\ref{table:ellipsoidparms}) and can see the star at optical
wavelengths.  In the IR, there is an excess arising from the
circumstellar disk, which reprocesses the stellar radiation.  Note
that, for pole-on viewing, we see the silicate features in emission,
since the disk is optically thin in the vertical direction.  As we go
to higher viewing angles, the optical depth to the central star
increases, and as a result, the star becomes more extincted in the
optical region.  Note that at almost edge-on viewing, a ``shoulder''
appears around $1\mu$m for the optically thick case.  This arises due
to the dominant effects of scattering the stellar radiation at these
wavelengths.  These scattering shoulders are also present in the
axisymmetric calculations presented by Efstathiou \& Rowan-Robinson
(1991), Sonnhalter et al. (1995), Men'schikov \& Henning (1997), and
D'Alessio et al. (1999).  At wavelengths longer than about $1\mu$m, the
albedo begins to drop rapidly (see Fig.~\ref{fig:dustopac}), and the
disk thermal emission begins to dominate, so the shoulder terminates.
Beyond $30\mu$m, the envelope becomes optically thin, so the spectrum
is independent of inclination and is dominated by the dust emission.

\placefigure{fig:diskSED}
\placefigure{fig:diskTemp}
\notetoeditor{Please place Figs.~\ref{fig:diskSED} \& 
\ref{fig:diskTemp} side by side.}

The two dimensional temperature structure for the disk-like models is
displayed in Figure~4b.  We see that at the inner edge of the envelope
($R_{\rm dust}$) there is little variation of the temperature
with latitude, while at large radii there is a clear latitudinal
temperature gradient, with the dust in the denser equatorial regions
being cooler than dust at high latitudes.

In the polar region, the material is optically thin to the stellar
radiation, so it heats up to the optically thin radiative equilibrium
temperature.  This temperature has a power law behavior, $T \propto 
r^{-0.4}$ for dust opacity ($\kappa \propto \lambda^{-1}$).  As can 
be seen in Figure~\ref{fig:diskTemp}, the polar temperature does
indeed have a power law decrease with a slope of approximately $-0.4$.

In contrast, the disk only displays this power law behavior at large
radii.  At the inner edge of the disk, the disk sees the same mean
(stellar) intensity as is present in the polar region ($J=WB$, where
$W=0.5\{1-[1-(R/r)^2]^{1/2}\}$ is the dilution factor).  Consequently,
the inner edge of the disk heats up to the optically thin radiative
equilibrium temperature --- the same as the polar temperature.  However
at larger radii, the opaque material in the inner regions of the disk
shields the outer regions from direct heating by the stellar
radiation.  This shielding reduces the mean intensity ($J<WB$).  As a
result, the outer disk is only heated to a fraction of the optically
thin radiative equilibrium temperature.  Thus the equatorial region is
cooler than the polar region.  Eventually at large enough radii, the
disk becomes optically thin to the heating radiation.  At that point,
it sees a radially streaming radiation field from an effective
photosphere that has a much lower temperature than the star.  From that
point outward, the disk temperature displays an optically thin power
law decrease with a slope that parallels the polar temperature.

The spherically symmetric calculation overestimates both the emergent
flux and the disk temperature.  In the optically thin limit, the IR
continuum is proportional to the mass of the circumstellar material, so
one would expect that a spherically equivalent mass would reproduce the
long wavelength spectrum.  Recall however that we can see the star at
pole-on viewing angles.  This implies that the disk does not reprocess
the entire bolometric luminosity of the star.  Consequently, the IR
excess is less than that in the spherical model.  Note that one can
nonetheless reproduce the edge-on SED using a spherical model if we
allow the density power law to depart from $1/r^2$ and change the size
of the spherical envelope.  This also was noted by Sonnhalter et al.
(1997), who found that they could fit the IR continuum of their disk
models by changing the radial dependence of the circumstellar density
and the outer radius for the same total envelope mass.

\begin{figure}
\epsscale{0.8}
\plotone{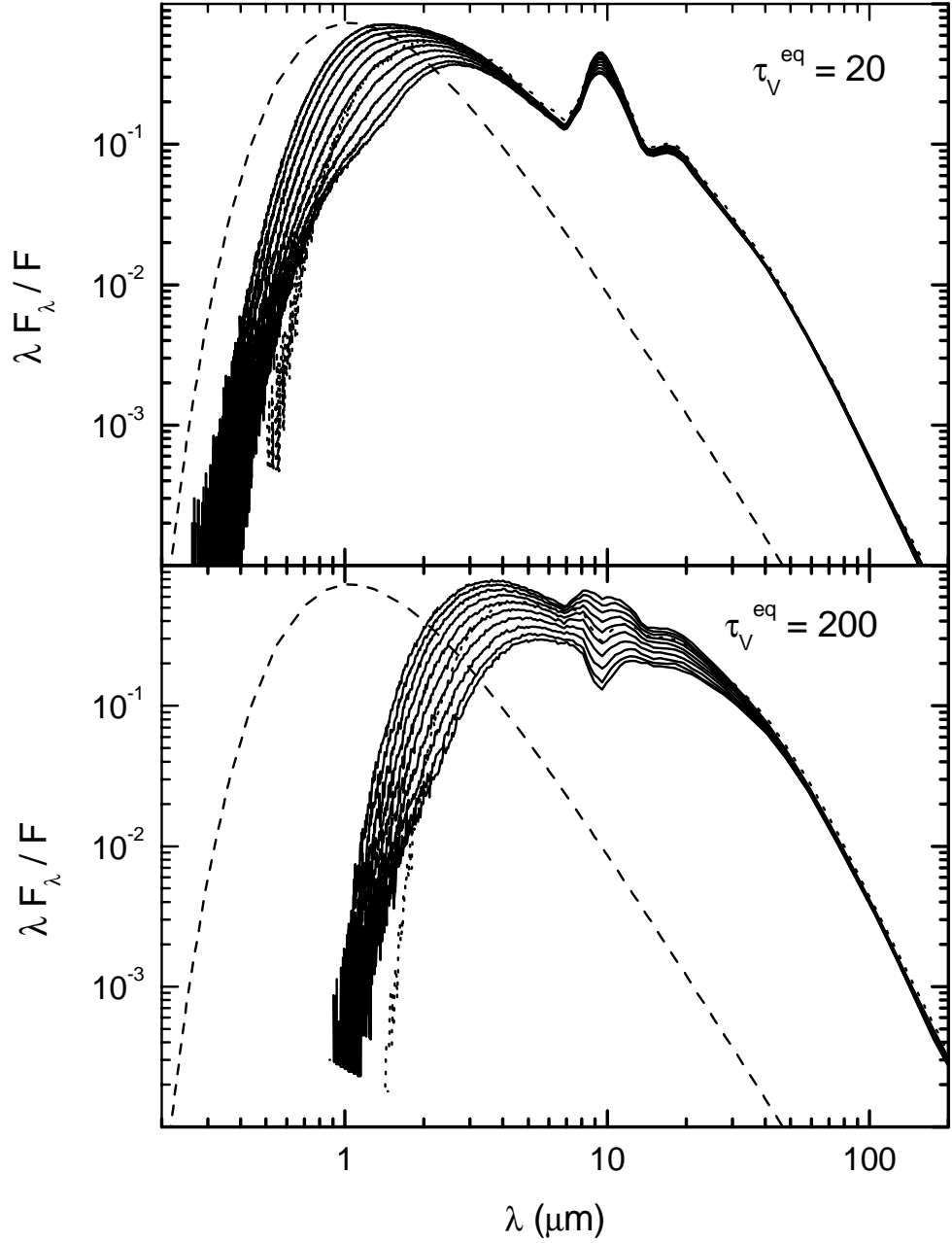}
\figurenum{5a}
\caption[] {Envelope Model SEDs.  Same as Fig.~4a, but for the envelope 
density distribution.}
\label{fig:envelopeSED}
\end{figure}

\begin{figure}
\epsscale{0.8}
\plotone{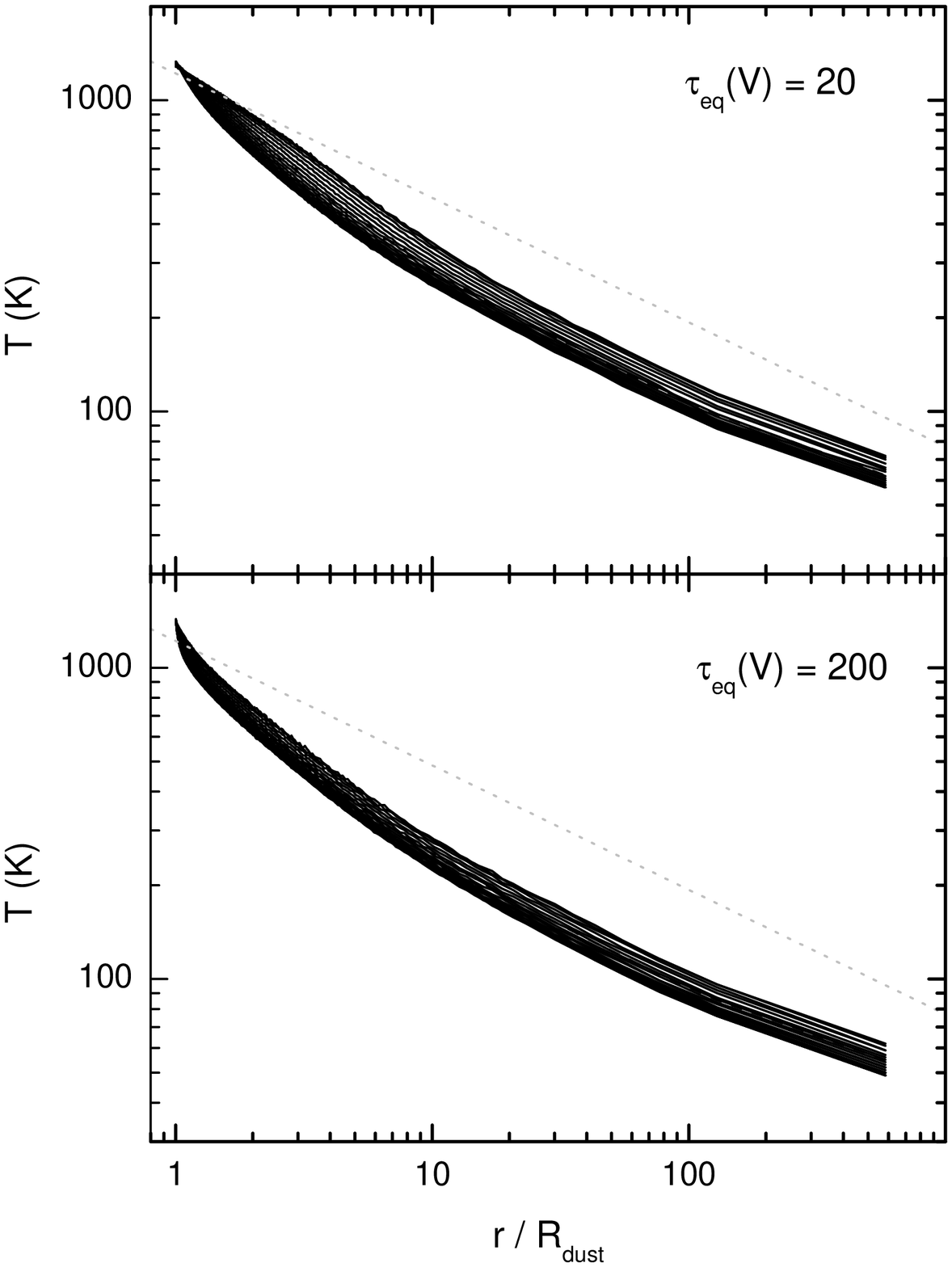}
\figurenum{5b}
\caption[]{Envelope Model Temperatures.  Same as Fig.~4b, but for the 
envelope density distribution.}
\label{fig:envelopeTemp}
\end{figure}

\subsection{Envelope Model}

The spectral energy distribution and temperature structure for the
oblate envelope are shown in Figures~\ref{fig:envelopeSED} and
\ref{fig:envelopeTemp}.  The optically thin envelope displays
significant extinction of the star at all viewing angles.  Close to
edge-on, scattering shoulders appear around $1\mu$m that are similar to
those seen in the disk model (Fig.~\ref{fig:diskSED}). Finally,
because the model is optically thin in the mid-IR, the silicate feature
is always in emission, and the SED is independent of viewing angle for
wavelengths longer than a few microns.

\placefigure{fig:envelopeSED}
\placefigure{fig:envelopeTemp}
\notetoeditor{Please place Figs.~\ref{fig:envelopeSED} \& 
\ref{fig:envelopeTemp} side by side.}

For the denser envelope, the star is extremely faint at optical
wavelengths.  Along with the increased extinction in the optical, the
scattering shoulders are less prominent, due to the thermal emission by
the envelope.  In the mid-IR, the envelope is optically thick edge-on
and optically thin pole-on.  Consequently, the silicate features go
from absorption to emission as the viewing angle changes from edge-on
to pole-on.  The general shape of the SEDs, which now peak in the
mid-IR for all inclinations, are reminiscent of spectra from embedded
(Class~I) T~Tauri stars ,which are commonly modeled using flattened
axisymmetric dusty envelopes (e.g., Adams, Lada, \& Shu 1987; Kenyon,
Calvet, \& Hartmann 1993; Men'shchikov \& Henning 1997; D'Alessio,
Calvet, \& Hartmann 1997).  Of these T~Tauri simulations, only
Efstathiou \& Rowan-Robinson (1991) have performed an exact calculation
for the SED and circumstellar temperature for the Terebey, Shu, \&
Cassen (1984) collapse model.

The temperature structure for our envelope models
(Fig.~\ref{fig:envelopeTemp}) is qualitatively similar to the
temperature structure of the disk models (Fig.~\ref{fig:diskTemp}).
The temperature at the inner edge of the envelope is independent of
latitude, while at larger radii the equatorial regions are cooler than
the polar regions.  The primary difference is that the latitudinal
temperature gradient is not as extreme.

Finally, we note that the equivalent spherically symmetric model better
reproduces the far IR spectrum of the envelope models than the disk
models.  This is because the envelope models reprocess the entire
bolometric luminosity, while in disk models, some of the stellar
luminosity escapes through the polar region.

\section{Discussion} 

We have developed a temperature correction procedure for use in Monte
Carlo radiation transfer codes.  We have tested our method against
other spherically symmetric benchmark codes and successfully matched
their results.  After verifying our method, we applied it to obtain
sample temperature distributions and SEDs for 2-D axisymmetric
disk-like models and mildly oblate envelopes.  These simulations
illustrate the important role envelope geometry can play when
interpreting the SEDs of embedded sources.

The primary limitation of our temperature correction procedure is that 
it only applies if the opacity is independent of temperature.  This is
not true for free-free opacity or hydrogen bound-free opacity, but 
it is true for dust opacity, which is the dominant opacity source
in many astrophysical situations.  The reason our method is limited to 
temperature-independent opacities is that two associated problems 
occur if the opacity varies when the cell's temperature changes.  The 
first is that the cell will have absorbed either too many or too few 
of the previous packets passing through the cell.  The second is the
associated change in the interaction locations of the previous packets,
which implies that the paths of the previous photon packets should 
have been different.  These problems do not occur if the opacity is
independent of temperature.

Lucy (1999) has proposed a slightly different method to calculate the
equilibrium temperature.  Instead of sampling photon absorption, he
directly samples the photon density (equivalent to the mean intensity)
by summing the path length of all packets that pass through a cell.
Typically more packets pass through a cell than are absorbed in the
cell, so this method potentially produces a more accurate measurement
of the temperature for a given total number of packets, especially when
the envelope is very optically thin.  The disadvantage of his method is
that it requires iteration to determine the envelope temperature.  In
principle one could partially combine both methods.  First, run our
simulation, adding pathlength counters to each cell.  After running all
packets, use the pathlength information to calculate a final
temperature.  This will provide a more accurate temperature that can be
used to calculate the source function.  After obtaining the source
function for each cell, it is a simple matter to integrate the transfer
equation to obtain the SED.

Another limitation of the Monte Carlo method is that it is not well
suited to envelopes with very high optical depths ($\tau > 100$ --
1000), unless there is an escape channel for the photons.  For example,
geometrically thin disks can be extremely optically thick in the radial
direction but optically thin in the polar direction, providing an
escape channel for the photons.  Similarly, dense envelopes can be very
optically thick to the illumination source, but optically thin to the
reprocessed radiation, which is a another escape channel.  In the event
no such escape channels exist, one must turn to other methods.  We are
currently investigating how to couple Monte Carlo simulation in the
optically thinner regions with other methods in the optically thick
interior.

The focus of the present paper has been on the development and
implementation of the temperature correction procedure.  For this
reason, we have made several simplifying assumptions regarding the
circumstellar opacity and geometry.  For example, we have assumed a
single temperature for all dust grains regardless of their size and
composition.  This differs from other investigations where different
types of grains can have different temperatures at the same spatial
location (e.g., the spherically symmetric radiation transfer code
developed by Wolfire \& Cassinelli 1986).  Similarly, we have not
implemented a procedure to solve for the location of the dust
destruction radius, which will be different for grains of differing
size and composition.  These issues are currently under investigation.

With the speed of today's computers, Monte Carlo radiation transfer
simulations can be performed in a reasonably short time.  The 2-D
simulations presented in this paper employed $10^8$ packets, requiring
about two hours of CPU time; the spherically symmetric cases used
$10^6$ packets, requiring about one minute.  So for continuum transfer,
Monte Carlo simulation is proving to be a very powerful technique for
investigating arbitrary density structures and illuminations.

\acknowledgements
We would like to thank Barbara Whitney and M{\'a}rio Magalh{\~a}es for 
many discussions relating to this work.  An anonymous referee also provided
thoughtful comments which helped improved the clarity of our presentation.  
This work has been funded by NASA grants NAG5-3248, NAG5-6039, and NSF grant 
AST-9819928.

\vfill\eject
\end{document}